  \documentclass[a4paper,11pt]{article}
\usepackage{amsmath,graphicx,color,mathrsfs,multicol}
 
\usepackage{verbatim}
 
\input epsf.sty

\newcommand{\com}[1]{}

\def\circa#1{\,\raise.3ex\hbox{$#1$\kern-.75em\lower1ex\hbox{$\sim$}}\,}
\makeatletter

\def\art{\@ifnextchar[{\eart}{\oart}}
\def\eart[#1]#2#3#4#5#6{{\rm #2}, {\em #3  #4} {\rm (#6) #5} ({\em #1})}
\def\hepart[#1]#2{{\rm #2, \em#1}}
\newcommand{\oart}[5]{{\rm #1}, {\em #2  #3} {\rm (#5) #4}}

\newcounter{alphaequation}[equation]
\def\thealphaequation{\theequation\hbox to
0.6em{\hfil\alph{alphaequation}\hfil}}
\def\eqnsystem#1{
\def\@eqnnum{{\rm (\thealphaequation)}}
\def\@@eqncr{\let\@tempa\relax \ifcase\@eqcnt \def\@tempa{& & &} \or
  \def\@tempa{& &}\or \def\@tempa{&}\fi\@tempa
  \if@eqnsw\@eqnnum\refstepcounter{alphaequation}\fi
\global\@eqnswtrue\global\@eqcnt=0\cr}
\refstepcounter{equation} \let\@currentlabel\theequation \def\@tempb{#1}
\ifx\@tempb\empty\else\label{#1}\fi
\refstepcounter{alphaequation}
\let\@currentlabel\thealphaequation
\global\@eqnswtrue\global\@eqcnt=0 \tabskip\@centering\let\\=\@eqncr
$$\halign to \displaywidth\bgroup \@eqnsel\hskip\@centering
$\displaystyle\tabskip\z@{##}$&\global\@eqcnt\@ne
\hskip2\arraycolsep\hfil${##}$\hfil& \global\@eqcnt\tw@\hskip2\arraycolsep
$\displaystyle\tabskip\z@{##}$\hfil
\tabskip\@centering&\llap{##}\tabskip\z@\cr}
\def\endeqnsystem{\@@eqncr\egroup$$\global\@ignoretrue} \makeatother

\usepackage{epsfig,multicol}
\usepackage{amsmath,amsfonts,amssymb}
\input epsf.sty

\usepackage{graphicx}
\usepackage[body={17.5cm, 22cm},right=2.2cm]{geometry}
\usepackage{amssymb}
\usepackage{amsmath}

\allowdisplaybreaks[1]
\def\o+{\oplus}

\def\beqa{\begin{eqnarray}}
\def\eeqa{\end{eqnarray}}

\def\zu{{\cal Z}_1}
\def\zd{{\cal Z}_2}

\newcommand{\be}{\begin{equation}}
\newcommand{\ee}{\end{equation}}
\newcommand{\bea}{\begin{eqnarray}}
\newcommand{\eea}{\end{eqnarray}}
\newcommand{\bg}{\begin{gather}}
\newcommand{\eg}{\end{gather}}
\newcommand{\bseq}{\begin{subequations}}
\newcommand{\eseq}{\end{subequations}}

\def\be{\begin{equation}}
\def\EQ{\begin{equation}}
\def\ee{\end{equation}}
\def\EN{\end{equation}}
\def\bea{\begin{eqnarray}}
\def\beq{\begin{eqnarray}}
\def\ba{\begin{eqnarray}}
\def\eea{\end{eqnarray}}
\def\ena{\end{eqnarray}}
\def\eeq{\end{eqnarray}}
\def\ea{\end{eqnarray}}
\def\bei{\begin{itemize}}
\def\eei{\end{itemize}}
\def\bee{\begin{enumerate}}
\def\eee{\end{enumerate}}

\def\lx{\left}
\def\rx{\right}

\def\a{\alpha}

\begin{document}
%\begin{flushright}
%{arXiv:0709...[hep-th]} \\
%%\\other numbers
%\end{flushright}
\vspace{1.5cm}

\begin{center}
{\LARGE \bf Induced gravity 
  on intersecting brane-worlds
  
  \vskip 0.4cm
  
  {\Large Part II:
  Cosmology}
}\\[1cm]

{
{\large\bf Olindo Corradini$^{a\,}$,\,  Kazuya Koyama$^{b\,}$
\,and\, Gianmassimo Tasinato$^{c\,}$
}
}
\\[7mm]
{\it $^a$ Dipartimento di Fisica, Universit\`a di Bologna and INFN Sezione
  di Bologna\\ Via Irnerio, 46 - Bologna I-40126, Italy} \\ 
 {E-mail}:  corradini@bo.infn.it\\[3mm]
{\it $^b$ Institute of Cosmology and Gravitation, 
University of Portsmouth  \\ Portsmouth PO1 2EG, UK 
}\\ {E-mail}: Kazuya.Koyama@port.ac.uk  \\[3mm]
{\it $^c$ Instituto de Fisica Teorica, UAM/CSIC  \\ Facultad de Ciencias C-XVI, C.U. Cantoblanco,
E-28049-Madrid, Spain} \\
 {E-mail}:  gianmassimo.tasinato@uam.es
\\[1cm]
\vspace{-0.3cm}

\vspace{1cm}

{\large\bf Abstract}

\end{center}
\begin{quote}

{
We explore cosmology of intersecting braneworlds with induced gravity
on the branes. We find the cosmological equations that control the evolution
of a moving codimension one brane and a codimension two brane that sits at the 
intersection. We study the Friedmann equation at the intersection, 
finding new contributions from the six dimensional bulk. These higher dimensional
contributions allow us to find new examples of self-accelerating configurations
for the codimension two brane at the intersection and we discuss their features.}

\end{quote}

\newpage
\tableofcontents

\bigskip

\bigskip

\bigskip

\section{Introduction}\label{intro}

%\begin{multicols}{2}
Brane-world models offer new perspectives for explaining the present day 
acceleration in purely geometrical terms, without the need to introduce dark 
energy \cite{dgp1,deffayetcosmology,deffayetdvali} (for a review see 
\cite{KK-review}). A celebrated example is the Dvali-Gabadadze-Porrati (DGP) 
model in a 5d spacetime~\cite{dgp1}. The brane action includes a quantum-induced Einstein-Hilbert action that recovers 4d gravity on small scales. 
This model realizes a so-called  {\it self-accelerating} solution that 
features a 4d de Sitter phase even though the 3-brane is
completely empty. However, so far, only codimension-one examples of such
solutions have been proposed and these backgrounds are known to
suffer from  ghost instabilities~\cite{Luty:2003vm}. An interesting
possibility then is to look for other such solutions in higher codimensional 
set-ups, initially introduced to address the cosmological
constant problem 
\cite{firstattemptscod2,carroll,sled,burgess1,Gregory:2000jc,dvaligabadadze}. 
This might lead to ghost free models \cite{newdvali}
(see however \cite{6d-induced1}).    

In this paper, as a continuation of \cite{ckt1}, we consider a codimension
two brane that lies at the intersection of two codimension one branes embedded 
in a six dimensional space. This system was studied in the past in the context 
of standard gravity \cite{int-standard} (cosmological properties were investigated
in~\cite{Cline:1999tq}), and Gauss-Bonnet gravity \cite{leetasinato} elaborating an idea developed in~\cite{lee2}. The latter was  generalized to higher-codimensional 
models in \cite{ignaciomulti}. Models with a generic angle between two 
intersecting branes were first considered in~\cite{Csaki:1999mz} and then further generalised into the so-called Origami-world in~\cite{origami}.  
More recently, in \cite{ckt1}, we added brane induced gravity terms to this
system to analyse the features of a configuration of static branes 
embedded in a time dependent, maximally symmetric background. We showed the
existence of new self-accelerating solutions, and of configurations with 
potentially interesting self-tuning properties.

In the present paper, we continue the analysis of this system by 
studying cosmological models, obtained by the motion of one of the branes
through the bulk, in a mirage approach \cite{kiritsis}. The energy momentum tensor 
different from pure tension on the branes causes the brane to move
and bend in the bulk, and induces cosmological evolutions from the point of 
view of observers sitting on the branes. We allow the branes to intersect at 
an arbitrary angle and to deform in the preferred shape.

The analysis of gravitational \cite{cod2-2} and cosmological \cite{cosmocod2} 
aspects of codimension two brane-worlds is a subject that is receiving some 
attention. Cosmology is mainly studied in the context of a mirage approach.
In higher codimensional brane-worlds, the mirage approach has
usually some drawbacks (critically examined, for example, in the introduction
of \cite{zavala}), mainly due to fine-tuning relations that the brane energy 
momentum tensors must satisfy. These are usually associated with the fact that 
an analogue of Birkhoff theorem does not hold in this case, in contrast to the 
codimension one case. In codimension one case, this theorem ensures that a system 
composed by a homogeneous and isotropic brane, moving through a static higher 
dimensional space, fully catches all the relevant time dependence of the system  \cite{ruthbirkhoff}. In our case, this is not true: generically, a moving higher
co-dimensional brane induces time-dependent effects in the bulk 
\cite{leetasinato,greeks}. In order to avoid the time dependence in the bulk, 
one must {\it impose} a static ansatz for the bulk geometry, and this is reflected 
on fine-tuning relations between matter on the brane and in the bulk.  
Nevertheless, it remains the most direct approach to study cosmological aspects 
of these models analytically.

The most interesting problem in this system is the isolation of the six
dimensional effects in the induced Friedmann equation on the codimension two 
intersection. As we will see, the Friedmann equation at the intersection 
receives contributions due to induced gravity terms on it which ensure the 
recovery of normal 4d cosmology in the relevant regimes. Moreover, there
are terms coming from induced gravity on the codimension one branes, of
the typical DGP form  \cite{deffayetcosmology}. Finally, and more interestingly
in our framework, the Friedmann equation contains also contributions that come 
from the six dimensional bulk. They vanish in the limit in which the branes
intersect at a right angle, but for generic brane configurations they
play an important role for the cosmological evolutions. Indeed, they can provide 
the late time acceleration, regardless of the energy content of the codimension 
two brane, generalising the self-accelerating branch of the codimension one DGP 
model to higher codimensions \cite{deffayetcosmology}. This fact 
has been realized already in \cite{ckt1}, but the present analysis is more general 
because we do not impose the maximal symmetry on the branes under consideration. 
By properly choosing the embedding for the codimension one branes, the six 
dimensional effects at the intersection can depend on the inverse of the 
induced Hubble parameter, and we will analyse the consequences of this in our discussion.
Another peculiar feature of our construction is that six dimensional contributions to 
the Friedmann equation are also associated to the non-conservation of the 
energy density at the intersection. During the cosmological evolution, the energy 
density indeed flows from the codimension one to the codimension two branes, unless
the codimension one branes intersect with a right angle. 
 
This paper is organised as follow. In Section \ref{sect:gen-form}, we will present
the general formalism that is necessary to study cosmological properties of the
systems we are interested in.  In Section \ref{arbitraryangl}, we apply this
formalism to a particular embedding for the codimension one branes, and in Section
\ref{sec:cosmsol} we study in some detail cosmological solutions derived from 
this embedding. Then, in Section \ref{sec:cosmapp}, we study applications of 
these cosmological solutions to some interesting situations. 
We conclude in Section \ref{sec:conc}.
 
\section{The general formalism}\label{sect:gen-form}
\subsection{The model}
We consider a system of two intersecting codimension one branes 
embedded in a six dimensional space-time. They intersect on a four dimensional 
codimension two brane, where observers like us can be localised.
We take an Einstein-Hilbert action for gravity in the bulk  
and we allow for induced gravity terms on the codimension one branes,
as well as on the intersection. Besides gravity, we allow for 
a cosmological constant term in the bulk, $\Lambda_B$, and 
for additional fields localised on the branes described by general 
Lagrangians $L$'s. The general action takes the form 

\begin{eqnarray}\label{model}
S&=&\int_{\rm bulk} d^{6} x \sqrt{-g}\ \biggl(\frac{M_{6}^{4}}2 R-\Lambda_B\biggr)
+\sum_{i=1}^2 \int_{\Sigma_i} d^{5} x \sqrt{-g_{(i)}}\ \biggl(\frac{M_{5,\,i}^{3}}2
R_{(i)}+L_{(i)}\biggr)\nonumber\\    
&
%+\sum_{i\neq j} \int_{\Sigma_i\cap\Sigma_j} d^{2+N} x
%\sqrt{-g_{(ij)}}\ \biggl(\frac{M_{2+N}^{N}}2 
%R_{(ij)}+L_{(ij)}\biggr)+\cdots
+&\int_{\Sigma_{\cap}} d^{4} x \sqrt{-g_\cap}\ \biggl(\frac{M_{4}^{2}}2 R_\cap+L_\cap\biggr),
\end{eqnarray}
%\ee
%\end{widetext}
where $\Sigma_{\cap}
\equiv\bigcap_i \Sigma_i$ denotes a three-brane at the intersection between 
all codimension-one branes 
 $\Sigma_i$. We can have different 
fundamental scales in the different regions of the space, $M_{6}$, $M_{5,\,i}$, 
and $M_{4}$. The induced gravity terms could be generated, as it was proposed
in the original model, by quantum corrections from matter loops on the brane.
It is also interesting to note that induced curvature terms appear quite generically 
in junction conditions of higher codimension branes when considering natural 
generalisations of Einstein gravity \cite{puretension2, charmousis} 
as well as in string theory compactifications 
\cite{antoniadis}, orientifold models and intersecting D-brane
models~\cite{Kohlprath:2003pu}.

%In this paper, we focus on a configuration with {\it static}  
%codimension one branes. They are characterised by tensions $\Lambda_1$ and
%$\Lambda_2$, while the intersection has tension $\lambda$. 
%The branes intersect with a generic angle, along the lines of 
%Origami-world~\cite{origami}.
The six dimensional bulk is characterised by a maximally symmetric geometry
\bea
ds^2 &=& A^2(t,z^1,z^2)\lx(\eta_{\mu\nu} dx^\mu dx^\nu +\delta_{k h}dz^k 
dz^h\rx), \nonumber\\  
A(t,z^1,z^2) &=& \frac{1}{1+\bar{H}
 t+k_i z^i}~.\label{bulkgengeo}
\eea
The parameters $\bar{H}$ and $k_i$ appearing in the warp factor $A$ satisfy the 
following relation
\be \label{cobulk}
\frac{\Lambda_B}{10}\,=\,\bar{H}^2-
k_1^2-k_2^2,
\ee
in order to solve the Einstein equations in the bulk. 

We embed a moving and a static codimension one branes ($\Sigma_2$ and $\Sigma_1$
respectively) on the background given by (\ref{bulkgengeo}).
The moving brane $\Sigma_2$ is characterised by an embedding
\be
X_{(2)}^{M}\,=\,\left( t, \, \vec{x}_3,  \, {\cal Z}_1(\omega_1) ,  \,
{\cal Z}_2(t, \omega_1)   \right). 
\ee

Here, $w_1$ is an embedding coordinate. In the following, for simplicity, 
we will demand that the intersection with the other brane lies at the 
position $w_1=0$, and that 
the function  $ {\cal Z}_1$ does not depend on time.
 The vectors $V$ tangent to $\Sigma_2$ are given by 
(we introduce indices on the left of $V$: they indicate which brane we are 
talking about) 
\be
{}^{(2)} V^M_{(a)} = \frac{\partial X^M}{\partial x^a},\quad x^a=(t,x^i,w_1),
\ee
where ${}^{(2)}V^M_{(t)}$ is (proportional to) the velocity vector 
\be
{}^{(2)} V^M_{(t)} = (1,\,0^i,\,0,\,\dot {\cal Z}_2) = \dot X^M~.
\ee
The other four vectors are
\bea
{}^{(2)} V^M_{(i)} &=& (0,\delta^M_i,0,0),\\
{}^{(2)} V^M_{(w_1)}  &=& (0,\,0^i,\,{\cal Z}'_1,\,{\cal Z}'_2)= X'{}^M~.
\eea
The normal vector to the brane is thus given by the conditions
\be
n_M V^M_{(a)}=0,\quad \forall a~.
\ee

Orthogonality with respect to the $i$ vectors simply removes from $n_M$ all its
3-dimensional space-like components. Imposing orthogonality w.r.t. 
${}^{(2)}V^M_{(w_1)}$ we then find
\be
n^{(2)}_{M}\,=\,\frac{A}{{\cal N}}\,\left(
- \dot{{\cal
      Z}}_2 {\cal
      Z}_1'  ,\,\vec{0}_3,\,- {\cal
      Z}_2',\,{\cal
      Z}_1'
\right),
\ee
with
\be
{\cal N}\,\equiv\,
\sqrt{{\cal
      Z}_1'^2+{\cal
      Z}_2'^2- 
       \dot{{\cal
      Z}}_2^2 {\cal
      Z}_1'^2}\,\,. \label{defofN}
\ee
\smallskip

\noindent
Doing exactly the same steps for the static brane 
$\Sigma_1$, with embedding
\be
X_{(1)}^{M}\,=\,\left( t, \, \vec{x}_3,  \, 0,  \, z_2   \right)\,\,,
\ee
the vectors tangent to the brane, ${}^{(1)}V_{(a)}^{M}$, are immediate
to find.  And the normal is simply
\be
n^{(1)}_{M}\,=\, A \,\left(
0 ,\,\vec{0}_3,\,1,\,0 
\right)\,\,.
\ee

\subsection{An useful change of coordinates}\label{uschcoo}

Proceeding identically as in the static case \cite{ckt1, origami},
it is useful to change a frame in order to impose the $Z_2$ symmetry 
in the case of a general angle. We go to coordinates parallel to 
the branes $(z_1, \, z_2)\,\to\, (
\tilde{z}_1, \, \tilde{z}_2)$, where
$$d \tilde{z}^{k}\,\equiv\, {\bf n}^{(k)}\cdot d {\bf z}\,.$$ 
One obtains two two-vectors
${\bf l}_{(1)}$ and ${\bf l}_{(2)}$ parallel to the branes:
\be
{\bf l}_{(1)} \,=\,\frac{1}
{{\cal Z}_1'}\left( {\cal Z}_1',\, { \cal Z}_2'
\right), \quad 
{\bf l}_{(2)} \,=\,\frac{\cal N}{{\cal Z}_1'}\,\left( 0,\,1
      \right),
\ee
and $d{\bf z}\,=\, {\bf l}_{(k)}\, d \tilde{z}^{k}$.  
%Notice that we explicitly indicated the $sign$ functions in the definitions
%of the $l$'s. 
Then the components of the 
 vectors ${}^{(2)} V$ parallel to the moving brane become
  \be
{}^{(2)}  {\tilde V}^M_{({w}_1)}\,=\,\frac{\partial \tilde
  X^M_{(2)}}{\partial w_1}\,=\,\left(0,\,\vec{0}_3,\, {\cal Z}_1',\,0\right)\,,
 \ee
and
 \be
{}^{(2)} {\tilde V}_{({0})}\,=\,\left(1,
\,\vec{0}_3,\, 
0,\, \frac{{\cal  Z}_1'  {\cal \dot Z}_2
 }{{\cal N}}\, 
 \right)~.
 \ee
Notice that the consistency relation 
 $$
 \frac{\partial^2 \, X^M_{(2)} }{\partial t\, \partial w_1}
\,=\, \frac{\partial^2\,  X^M_{(2)} }{\partial w_1\, \partial t},
 $$
in our case implies the condition
 \be\label{con2impose}
 \frac{\partial}{\partial w_1}\,\left(
  \frac{{\cal  Z}_1'  {\cal \dot Z}_2 }{{\cal N}}
  \right) \,=\,0\,.
 \ee
The normal to the moving brane   becomes in these coordinates

\be
\tilde{n}^{(2)}_{M}\,=\,A \,{\cal T}\,\left(
-\frac{ \dot{{\cal Z}}_2 {\cal Z}_1'}{{\cal N}}
,\,\vec{0}_3,\,0,\,1
\right),
\ee
where ${\cal T}$ is a normalization factor that we will 
fix once we define the six dimensional metric. 

In order to proceed, 
 we must take into account that the branes are fixed points of $Z_2$ 
symmetries. We focus our analysis on the moving brane $\Sigma_2$
in order to compute Israel junction conditions at its position: the analysis 
for the static brane $\Sigma_1$ can be performed along similar lines.
The $Z_2$ symmetry acting on the static brane $\Sigma_1$ implies the invariance 
of the 6d metric under $\tilde z^1 \to -\tilde z^1$, that can be obtained
replacing $\tilde{z}^1 \to |\tilde{z}^1|$.

After the change of frame, imposing the $Z_2$ symmetry, the six dimensional 
metric becomes 
\bea
\tilde{\gamma}_{mn} &=&
\frac{1}{{\cal Z}_1'^2}
\left(
\begin{array}{ccc}
{\cal Z}_1'^2+{\cal Z}_2'^2
 \,\,& &{\cal N}\,{\cal Z}_2'
  \,sign{(\tilde{z}_1)} 
 \\ \\
 {\cal N} \,{\cal Z}_2'  
 \,sign{(\tilde{z}_1)} 
 & &{\cal N}^2
\end{array}
\right)~,
\eea
with inverse 
%~\footnote{There are subtleties here related to terms $sign^2(\tilde
 % z^1)$, for example
   %appearing in the determinant of the metric. We %discuss about this in
  %Appendix~\ref{appendix:singular}.}
\bea
\tilde{\gamma}^{mn} &=&
 %{\vec l}_{(m)} \cdot {\vec l}_{(n)}
 %\nonumber\\ 
%\,=\,
\frac{1}{{\cal N}^2\,{\cal C}^2}
\left(
\begin{array}{ccc}
{\cal N}^2\,\,
 & &-{\cal N}\,{\cal Z}_2' 
 \,sign{(\tilde{z}_1)} 
 \\ \\
 -{\cal N} \,{\cal Z}_2'  \,\,
 \,sign{(\tilde{z}_1)} 
 & &{\cal Z}_1'^2+{\cal Z}_2'^2
\end{array}
\right)\,\,,
\eea
%with ${\cal M}^2\equiv {\cal N}^2 \frac{{\cal Z}_1'^2+{\cal
  %  Z}_2'^2(1-sign^2(\tilde z^1))}{{\cal Z}_1'^2}$. 
  where 
  \be
  {\cal C}^2\,=\,1+\frac{{\cal Z}_2'^2}{{\cal Z}_1'^2}\,\left(
1-sign^2(\tilde{z}_1)\right).\label{defofC}
  \ee
  
With this information, we determine the normalization
factor ${\cal T}$ by requiring that the normal $\tilde{n}^{(2)}$ 
has a unit length. Then ${\cal T}$ is determined as 
\be
{\cal T}\,=\,\frac{{\cal C}\,{\cal N}}{{\cal Q}},  \label{defofT}
\ee
with ${\cal N}$ defined in formula (\ref{defofN}), 
 ${\cal C}$  in (\ref{defofC}),
 and 
\be
{\cal Q}^2\,=\,
{\cal Z}_1'^2+ {\cal Z}_2'^2\,-\,
{\cal C}^2\,
{\cal Z}_1'^2 \dot{{\cal Z}_2}^2. \label{defofQ}
\ee

\smallskip
On the other hand the induced metric is invariant under bulk reparametrisation, 
and thus reads
 \bea\label{genindmet2bis}
 d s_{5, \Sigma_2}^2=A^2\left(t, w_1\right)
 \,\biggl[ &-&\left( 1-\dot{\cal Z}_2^2  \right) d t^2
 + d \vec{x}_3^2+ \left( {\cal Z}_1'^2+  {\cal Z}_2'^2\right) 
 d w_1^2 \nonumber \\&+&2\,\dot{\cal Z}_2{\cal Z}_2'
 \,sign{(\tilde z_1)}\,dt\, dw_1
 \biggr]~.
 \eea
The inverse induced metric is given by
\bea
h^{ab} =A^{-2}\left(
\begin{array}{ll}
\delta^{ij} & 0\\
0& H^{\alpha\beta}
\end{array}\right),
\eea 
with
\bea
H^{\alpha\beta} ={\cal Q}^{-2}\left(
\begin{array}{ll}
-({\cal
  Z'}^2_1+{\cal Z'}^2_2) & \dot{\cal Z}_2{\cal Z'}_2\,sign{(\tilde z_1)}\\[2mm]
\dot{\cal Z}_2{\cal Z'}_2\,sign{(\tilde z_1)} & 
1 -\dot{\cal Z}^2_2
\end{array}\right),
\eea 
and ${\cal Q}$ defined in eq. (\ref{defofQ}).
 At the intersection, the four dimensional metric is given by
\begin{equation}
ds_{4} = A^2(t)[
-(1-\dot{\cal Z}^2_2 )dt^2 +d{\bf x}^2].
\end{equation}

\subsection{Extrinsic curvature}
Given all this information, one can compute the components of the extrinsic curvature
at the position of the brane $\Sigma_2$, using the general formula
\be
 K_{mn} = \tilde V^M_{(m)}\tilde V^N_{(n)} 
\tilde \nabla_M \tilde n_N~.
\ee  
Since the expression for $K_{mn}$ is invariant under bulk reparametrisation, 
to evaluate the right hand side of the previous expression one can use
the six dimensional metric in the original frame, or in the frame parallel to 
the brane.

The calculation of the regular part of $K_{mn}$ is easier to work out in the 
original frame. The non vanishing components are the following
\bea
&& K_{00} = -
\frac{A}{{\cal N}}
 \ddot{{\cal Z}}_2 {\cal Z}_1'
\,
-\,
\frac{A^2}{{\cal N}}\,\left(1- \dot{\zd}^2\right)
\,
{\cal K}(w,t)  \,,\label{expr00ec}
\\
&& K_{w_1 w_1} =
\frac{A}{{\cal N}}
\left( 
{\cal Z}''_1 {\cal Z}_2'- {\cal Z}''_2 {\cal Z}_1'
\right)\,
\,+\,
\frac{A^2}{{\cal N}}\,\left({\zu}'^2 +{\zd}'^2\right)
\,
{\cal K}(w,t)
\,,\\&& 
 K_{0 w_1} = \,-sign{({\tilde{z}}_1)} \frac{A }{{\cal N}}\,
  {\cal Z}_1' \dot{{\cal Z}}_2' 
\,+\,sign{(\tilde{z}_1)}\,
\frac{A^2}{{\cal N}} \,{\cal K}(w,t)
\,\dot{{\cal Z}}_2 {\cal Z}'_2
\,, \label{expr0w1ec}\\&&
K_{ij} =  \frac{A^2\delta_{ij}\,\,
{\cal K}(w,t)}{{\cal N}}
 \label{exprijec}\,.
 \eea    
where
\be 
{\cal K}(w,t)= k_1  {\cal Z}'_2-k_2
 {\cal Z}'_1- \bar{H} \dot{{\cal Z}}_2 {\cal Z}_1'
\ee
In addition, the component $ K_{w_1 w_1}$ of the extrinsic curvature may contain 
terms localised at the intersection due to the presence of the $sign$ functions 
in the six dimensional metric. Let us then look for the singular pieces of the 
extrinsic curvature
\bea
K_{ab}|_{sing} = \tilde{V}^M_{(a)}
 \tilde{V}^N_{(b)} \nabla_M n_N|_{sing}~. 
\eea
This quantity is much easier to calculate in the tilted reference frame. 
There are, a priori, two classes of contributions to the singular pieces. 
The first one comes from partial derivatives acting on $n_N$, due to the
$sign$ function included in ${\cal T}$. However such a contribution
is proportional to $n_N$ itself as the $sign$ function only appears in the
prefactor of $n_M$, and thus vanishes due to the orthogonality between $n_M$ 
and $V^M_{(a)}$.

 Then
\bea
K_{ab}|_{sing} =
-  V^M_{(a)}V^N_{(b)}\ n_R\ \Gamma_{MN}^R|_{sing},
\eea
and since $\Gamma^0_{MN}$ has no singular part, one is left with
$\Gamma^{\tilde z_2}_{MN}$. Its only singular component is 
$\Gamma^{\tilde z_2}_{\tilde z_1 \tilde z_1} = g^{22}\partial_1 g_{12}$
% = 
%2\frac{{\cal Z}_2'}{{\cal Z}_1'^2}\frac{{\cal Z}_1'^2+{\cal Z}_2'^2}{\cal N} 
%\delta(\tilde z_1)$. 
 since they involve derivatives of the $sign$ function. 
Then, in the end, we find that  
\be
K_{w_1 w_1}|_{sing} = -A\,{\cal K}\,
 \frac{{\cal Z}_1'^2+{\cal Z}_2'^2}{{\cal C}^2\,{\cal N}}
\, {\cal Z}_2'
% \frac{{\cal Z}_2'}{{\cal Z}_1'}
%\left({\cal Z}_1'^2+{\cal Z}_2'^2
%\right)
\, \left( 
\frac{\partial }{\partial \tilde{z}_1}\,sign(\tilde{z}_1)
\right),
\ee 
is the only singular component of the extrinsic curvature.
Notice that the previous expression 
contains  products of distributions, since the ${\cal C}$ contains
squares of $sign$ functions. 
 In order
to dispel any doubts about how to define
 such singular expression,
 it is convenient {\it not} to
 set 
 $ 
\frac{\partial \,sign(\tilde{z}_1) }{\partial \tilde{z}_1}
\,=\,2\frac{\delta(w_1)}{{\cal Z}_1'}
$, but instead  maintain
the derivative of the $sign$ function. Later, in
the specific examples we will discuss, in order
to extract the singular terms localized at the intersection we will perform 
 an explicit integration on a small interval
 around the singularity. The result of
 this integration will provide the value of the various quantities
 localized at the intersection.

\smallskip

% a straightforward calculation
%(see Appendix~\ref{appendix:extrinsic})
%shows
%that  the only singular component is 
% %(see Appendix~\ref{appendix:extrinsic})
%\begin{eqnarray}
%  K_{w_1}{}^{w_1}|_{\rm sing}&=& -\delta(w_1)\,
 %\frac{2 }{A\,{\cal N}^3}
 %\, \frac{ {\cal Z}_2' }{ {\cal Z}_1'} \, \left(
%1
%%- \dot{{\cal Z}}_1^2
%-  \dot{{\cal Z}}_2^2  \right)\,\left( 
 %{\cal Z}_1'^2+{\cal Z}_2'^2 
 %\right)
 % \end{eqnarray}
%that reduces, in the static case, to the results
%of \cite{ckt1}. 

For the static brane $\Sigma_1$, the extrinsic curvature is 
simply given by
\begin{equation}
K^{a}_b = -k_1 \delta^a_b.
\end{equation}

\subsection{Junction conditions}
The previous expressions for the extrinsic curvature
are important in order to obtain the  equations that govern the induced
cosmology on the brane. They are dictated by the Israel junction conditions
\begin{equation}
2\left[  \hat{K}_{a b}
 \right] \,\equiv\,  2\left[ K_{ab}-Kh_{ab} \right] \,
=\,-\frac{1}{M_{6}^4}({ S}_{ab}+S^{loc}_{ab}),\label{braneeq}
\end{equation}
where  $\left[ X\right] \equiv 
(X(\Sigma_{2,+})-X(\Sigma_{2,-}))/2$,  while the  induced 
  codimension one brane metric is $h_{ab}$.
The extrinsic curvature tensor evaluated on $\Sigma_2$  is given 
by $K_{ab}=h^M_{a} h^N_{b}\nabla_M n_N$ with $K=K^a\,_{a}$, and 
  energy momentum tensors relative
to matter localised on $\Sigma_2$ , appearing on the right hand side of
(\ref{braneeq}),  are calculated in the usual way:
\begin{eqnarray}
{ S}_{ab}&=&-\frac{2}{\sqrt{-h_{(2)}}}
\frac{\delta\left( \sqrt{-h_{(2)}}  {\cal L}_{(2)}\right)}{\delta h_{(2)}^{ab}}\,, \\
S^{loc}_{ab}&=&-\delta(\Sigma_1)\delta_a^\mu\delta_b^\nu
\frac{2}{\sqrt{-h_{(2)}}}\frac{\delta \left(\sqrt{-h_{}} 
{\cal L}_{\cap}\right)}{\delta h_{\cap}^{\mu\nu}}\nonumber\\ &&\equiv 
\delta(\Sigma_1) \sqrt{\frac{- h_{\cap}}{-h_{(2)}}}  \delta_a^\mu\delta_b^\nu 
S_{\mu\nu}\,\,.
\end{eqnarray}
In our model the localised energy-momentum tensor also includes  contributions
from the induced gravity terms. The last quantity $S^{loc}_{a b}$ denotes energy 
momentum tensor that is  localised on the intersection $\Sigma_{\cap}$ 
between the branes. Notice the presence of the factor $ \sqrt{ h_{\cap}/ h_{(2)}}$ 
that renders the expression covariant with respect to the metric at the 
intersection.
 
% We will suppose that the energy momentum tensor on the
 %branes acquire a perfect fluid form.  Then we can calculate
 %the contribution to the energy density
 %at the intersection, that  is due to the singular part
 %of the extrinsic curvature  

\bigskip

In the previous discussion, we learned that the only singular term of the 
extrinsic curvature for the brane $\Sigma_2$, that is localised at the intersection, 
is contained in $K_{w_1 w_1}$. This implies that the six dimensional contributions
to the energy momentum tensor must be proportional to the induced metric, 
$S^{loc}_{\mu \nu} \,=\, f(x^\mu) \,h_{\cap\,\mu \nu}$, for some function $f$.
We still do not know whether this function $f$ is a constant (in which case, it 
corresponds to a pure tension) or not, since we do not know whether the conservation of the 
energy holds at the intersection or not. The Codazzi equation holds in 
this case~\footnote{The RHS of this formula vanishes because
both the bulk and the static brane have maximal symmetry
\cite{shiromizu}.}
\be 
\nabla_{a} \hat{K}^{a}_{\,\, b}\,=\,0\,\hskip0.5cm \Rightarrow\hskip0.5cm
\nabla_{a} S^{a}_{\,\, b}\,=\,0,
\label{conservequa}
\ee
which means that there is no exchange of energy between the bulk and the 
codimension one branes. But the previous relations may contain singular terms, 
associated with an exchange of energy between the codimension one and the codimension two branes. This is indeed what generically happens, and we will encounter an example of 
this phenomenon in Section \ref{arbitraryangl}. Singular terms in the first of the
previous formulae can appear if $\hat{K}^a_b$ has singular pieces, or if some of 
its components become singular when covariant derivatives act on them. 
This possibility occurs when the angle between the brane is not right: then the 
component $K_0^{\,\,w_1}$ is non-vanishing at the intersection, and, being proportional 
to the $sign$ function (see eq. (\ref{expr0w1ec})),  it normally generates an 
additional singular term.

\section{Applications}\label{arbitraryangl}
We then consider a system with the static brane $\Sigma_1$, and the moving brane $\Sigma_2$. 
Here, $\Sigma_2$ is free to move and bend arbitrarily.
We assume that the induced gravity term on the moving codimension one brane vanishes: $M_{5,\,2}=0$ for simplicity. For this brane, we take an embedding
\be
 X^{M}\,=\,\left( t, \,\vec{x}_3, \,{\cal Z}_1(w_1)
 , \,{\cal Z}_2(t,w_1)
  \right),
\ee
with
\bea
{\cal Z}_1&=& w_1  \cos{\alpha(w_1)},\\
{\cal Z}_2&=& z_2(t,w_1) + w_1  \sin{\alpha(w_1)},
\eea
where we wrote the two functions ${\cal Z}_i$ in terms
of the auxiliary functions ${z}_2$ and $\alpha$.
We demand that these functions are continuous with respect to the variable $w_1$,
at the position of the intersection $w_1=0$, and to avoid subtleties
related with the reflection symmetry at the  
intersection we require smoothness conditions $z_2'(t,0)=z_2''(t,0)=
\alpha'(0) =0$
\footnote{Asking that only the first derivative vanish at the intersection may 
be enough to ensure sufficient smoothness to render the system well
behaved. In fact later we will briefly mention a situation where a non-vanishing second 
derivative $z_2''(t,0)\neq 0$ can turn out to be useful.}.
From these definitions, we have
\be
\begin{split}
& \dot {\cal Z}_1 = 0, \quad {\cal Z}_1' =  
\cos{\alpha} - w_1 \alpha' \sin{\alpha}, \\
& \dot {\cal Z}_2 = \dot z_2 , \quad 
 {\cal Z}_2' = z_2'+\sin{\alpha}+ w_1 \alpha' \cos{\alpha}.
\\
%& \dot {\cal Z}_1^2 +\dot {\cal Z}_1^2 = (\dot \alpha w_1 +\dot z_2
  %    \cos\alpha)^2 +\dot z_2^2 \sin^2\alpha
\end{split}
\ee
\smallskip
%Notice that the equation (\ref{con2impose}) imposes, for this embedding,
%\be\label{costancon}
 %\dot{\alpha} (w_1=0)\,=\,0\,.
% \ee
Notice that the previous
embedding satisfies the constraint (\ref{con2impose}); for a time-dependent angle  (\ref{con2impose})
would instead imply $\dot\alpha(t,\,w_1=0) =0$.

%that is
%automatically satisfied in our case since
%we take the angle $\alpha$ as
 %constant. 
%This is due to our choice of embedding and simplifies the calculations 
%when we focus on the properties of the intersection. 
 
The junction condition on the static co-dimension one brane gives the tension 
$\lambda_1$ as 
\be
\lambda_1 = 6 M_{5,1}^2 (\bar{H}^2-k_2^2) + 8 M_6^4 k_1.
\ee
The induced metric on the brane $\Sigma_2$ is obtained by plugging the previous 
expressions in (\ref{genindmet2bis}). The complete calculation of the cosmological 
behaviour for the moving codimension one brane is complicated as 
the brane is inhomogeneous, but it can be obtained straightforwardly from the general formulae (\ref{expr00ec})--(\ref{exprijec}). In the next subsection we discuss some 
of its properties that are useful when comparing them with cosmology on the codimension 
two brane.

\subsection{Cosmology on the moving four brane}
The cosmological evolution on the moving brane $\Sigma_2$ is complicated
by the fact that its induced scale factor and energy momentum tensor must be  
inhomogeneous, in order to satisfy Israel junction conditions
(\ref{expr00ec})--(\ref{exprijec}): such conditions require some
off-diagonal components of the energy momentum tensor $S^a{}_b$ to be
non vanishing. In what follows we will only need the explicit form of
the codimension-one equations evaluated at the intersection. We thus
concentrate on such a limit where the form of the needed energy 
momentum tensor is the following:
\bea
S^a{}_b =\left(
\begin{array}{ccccc}
-\rho_2 & 0& 0& 0& 0 \\
0&p_2 &0&0&0\\
0&0&p_2&0&0\\
0&0&0&p_2&0\\
\chi &0&0&0 &p_2
\end{array}
\right),\quad {\rm at}\ w_1=0~. 
\eea
Then the junction conditions impose the following relations
\bea
\rho_2 &=& -8M_6^4 {\cal N}^{-1} {\cal K},\\
p_2 &=& 8 M_6^4 ({\cal N}^{-1} {\cal K} +\frac{1}{4}{\cal N}^{-3} A^{-1} \ddot{z}_2 
\cos \alpha),\\
\chi &=& 2 M_6^4 A^{-1} {\cal N}^{-3} \dot{z}_2 \ddot{z}_2 
\sin \alpha \cos \alpha,
\eea 
where we define ${\cal N}=\sqrt{1- \cos^2 \alpha \dot{z}_2^2}$ and 
${\cal K}= k_1 \sin \alpha - (k_2 + \bar{H} \dot{z}_2) \cos \alpha$.
%This system of equations is enough to completely determine the cosmological 
%evolution of the codimension one branes. 

\subsection{Cosmology at the intersection}
We start from discussing the contributions from the brane $\Sigma_2$ to 
the codimension two brane. At the intersection, characterised by $w_1=0$, 
the induced metric is straightforwardly extracted from the five dimensional
one and is simply given by
\be
 d s_4^2 = A^2(0,t)\,\Big\{- \left[ 1- \dot{z}_2^2
 \right] d t^2 + d \vec{x}_3^2
  \Big\}= -d \tau^2 + a(\tau)^2 d \vec{x}_3^2, \quad 
  d \tau = A(0,t) \sqrt{1-\dot{z}_2^2} dt.
 \ee
Then the induced Hubble parameter is 
\begin{equation}
H = \frac{1}{a(\tau)} \frac{d a(\tau)}{d \tau} =
\frac{\bar{H}+k_2 \dot{z}_2}{\sqrt{1- \dot{z}_2^2}}.
\label{genHin}
\end{equation}
In order to find the Friedmann equation at the intersection,
we have to extract the singular part of the Israel junction conditions for the 
codimension one branes. This singular part receives contributions from
the energy momentum tensor localised on the codimension two brane
(containing also the induced gravity terms at the intersection),
from the induced gravity terms on the codimension one branes, and from
singular contributions of the extrinsic curvature terms. 
The final contribution represents the most interesting feature of our model
since it corresponds to six dimensional contributions to four dimensional
physics. We start our discussion with their evaluation.

Recall that only singular part on the extrinsic curvature
for brane $\Sigma_2$ is contained in the $(w_1,w_1)$ component.
From its expression one straightforwardly obtains
\be
\hat{K}_{\mu}^{\,\,\nu\,(sing)}\,=\,
 \frac{{\cal T}\,  \sin{\alpha}\,}{A\,{\cal N}\,{\cal Q}^2\,{\cal C}^2}\,
  \left( 1-\dot{z}_2^2 \right)
  \,\left(\frac{\partial}{\partial 
\tilde{z}_1} sign(\tilde{z}_1)  \right)\,\delta_{\mu}^{\nu},
\ee

%\be\label{sinpar}
%\left(K_{w_1}^{w_1}\right)^{sing}\,=\, -\frac{2}{A\,\sqrt{1-\cos^{2}{\alpha}
%\,\dot{z}^2_2}}
 %\,
 %\frac{1-
%\dot{z}^2_2}{1-\cos^{2}{\alpha}\,
%\dot{z}^2_2}
 %\,\,\frac{\sin{\alpha}}{\cos{\alpha}}
 %\,
%\delta(w_1),
%\ee
%so that $(\hat K_{mn})^{sing } = -g^{(5)}_{mn} \left(K_{w_1}^{w_1}\right)^{sing}$.
\noindent
as the only singular piece, and it is easy to see that
 its contribution
to the  energy momentum tensor at the intersection 
results proportional to the induced metric
 (recall the definitions of ${\cal T}$, 
 ${\cal N}$, ${\cal C}$ and ${\cal Q}$ 
 respectively in eqs. (\ref{defofT}),   (\ref{defofN}),
  (\ref{defofC}) and  (\ref{defofQ})). 
 Let us, for example, calculate its contribution to the
 energy density at the intersection. Plugging
 the previous expressions inside the Israel
 equations, one finds the
 relation
 
 \be
\frac{\delta \rho}{4 M_6^4}\,\left(\frac{\partial}{\partial 
\tilde{z}_1} sign(\tilde{z}_1)  \right)\,
=\,
 \frac{
 {\cal T}\,\tan{\alpha}
 }{{\cal C}^2{\cal N}\,{\cal Q}}\,
  \sqrt{ 1-\dot{z}^2 }
  \,\left(\frac{\partial}{\partial 
\tilde{z}_1} sign(\tilde{z}_1)  \right)\, ,\label{masexp}
\ee
where $\delta \rho$ indicates the contribution
to the brane energy density, due to purely six dimensional
effects.

The previous equation contains $sign^2(\tilde{z}_1)$
functions inside the expressions for ${\cal N}$,
${\cal Q}$, and $\cal C$. The safest way to handle
them is to integrate both sides of (\ref{masexp})
 along an infinitesimally
  small interval centered at the origin, furnishing the
  value of the contribution to energy density localized
  at that point.
Performing the integration  we get the expression

\be
\frac{\delta \rho}{2 M_6^4}\,=\,
\,\tan{\alpha}
 % \frac{{\cal Z}_2'}{{\cal Z}_1'}\,
  \sqrt{ 1-\dot{z}_2^2 }
  \,
  \int_{-1}^{+1}\,\frac{d \,sign(\tilde{z}_1)}{{\cal C} \,{\cal Q}^2}.
\ee

After a few calculations, the previous expression can be written as 
\bea
\frac{\delta \rho}{2M_6^4}&=&
\frac{\sin{\alpha}}{ \sqrt{ 1-\dot{z}_2^2 }}
%\left({\cal Z}_1'^2+{\cal Z}_2'^2\right)
%\,
 %\frac{{\cal Z}_2'}{\sqrt{{\cal Z}_1'^2+{\cal Z}_2'^2}}
  \,
  \int_{-1}^{+1}\,d \,x\,\frac{1}{\sqrt{1-\sin^2{\alpha}\,x^2}}\,\frac{1}{1+\frac{\dot{{z}}^2_2 \sin^2{\alpha}}{ 1-\dot{z}_2^2} x^2}.
\eea
By performing the integration, we get
\be
\delta\rho\,=\,4 M_6^4\,\arctan{\left[\frac{\tan{\alpha}}{\sqrt{1-\dot{ z}_2^2}}\right]}.
\ee
Notice that, in the limit of static brane $\dot{z}_2=0$,
this contribution is proportional to the angle $\alpha$,
 exactly as happens in the case of codimension
 two conical singularities.

%Then one obtains the following contribution to the energy momentum 
%at the four-dimensional intersection:
%\be\label{eninint}
%T_{\mu \nu}^{loc}\,=2M_{6}^4 \delta_\mu^m
%\delta_\nu^n~\frac{\sqrt{h_{(2)}}}{\sqrt{h_{\cap}}} ~\left(\hat
%K_{mn}\right)^{sing }=\,4M_{6}^4\,
%\frac{\sqrt{1-
%\dot{z}^2_2}}{1-\cos^{2}{\alpha}
%\,\dot{z}^2_2}
% \,\,\frac{\sin{\alpha}}{\cos{\alpha}}
% \,
%\,h_{\cap \;\mu\nu}.
%\ee
%As expected, it is proportional to the induced metric at the intersection. 

\bigskip

Proceeding with our calculation, we can determine the contributions to the 
intersection from the induced gravity terms on $\Sigma_1$ (recall that we have 
chosen $M_{5,\,2}=0$ so there are no induced gravity terms on the moving brane 
$\Sigma_2$). 
We find 
\be
\left(G_0^0\right)^{sing}\,=\,-\frac{6 (k_2+\bar{H} \dot{z}_2)}
{\sqrt{1-\dot{z}_2^2}}.
\ee

Putting all this information together, we find the following 
equation  relating energy density
to geometrical
quantities (here $\rho$ indicates the total
energy density at the intersection)
\bea\label{Hubtdangle}
\rho &=& 3 M_4^2 H^2
%\nonumber\\ &+& 6 
+ 6  M_5^3 \,
%\sqrt{\frac{1-
%\dot{z}^2_2}{1-\cos^{2}{\alpha}\,
%\dot{z}^2_2}}\left(k_1 \cos{\alpha}+k_2 \sin{\alpha}
%+k_2 \frac{\sin{\alpha}\,\dot{z}_2^2}{1-
%\dot{z}^2_2}\right)
%+
\frac{k_2+\bar{H} \dot{z}_2}{\sqrt{1-\dot{z}_2^2}} 
%\nonumber\\
%&+&%\frac{
+4 M_6^4\,\arctan{\left(\frac{\tan{\alpha}}{\sqrt{1-\dot{ z}_2^2}}\right)},
% \,\,\frac{\sin{\alpha}}{\cos{\alpha}}
%}
\eea
that can be interpreted as the Friedmann equation
for an observer localized at the intersection.
The induced gravity terms on the codimension one branes do not
induce a violation of the continuity equation at the intersection
because the intersection can be seen as codimension one object from
the point of view of the four branes and then the properties
of the Israel formalism for the codimension one brane ensure 
the conservation of energy (see eq. (\ref{conservequa})). 
On the other hand, the last, six
dimensional term in Eq.~(\ref{Hubtdangle}) is explicitly time dependent, 
while we know that it appears as a tension term in the effective energy
momentum tensor at the intersection. This is because it is proportional
to the induced metric. This indicates that this 
term is likely to be associated with a violation of the continuity
equation at the intersection.

This issue can be understood by re-considering the Codazzi equation:
 \bea
 \nabla_M\,\hat{K}_N^{M}&=&\left( \nabla_M\,\hat{K}_N^{M}\right)^{\rm (reg)}
 +
 \left( \nabla_M\,\hat{K}_N^{M}\right)^{\rm (sing)}\,=\,0,
 \label{genfornocont}
 % \hskip0.3cm
 %\Rightarrow \hskip0.3cm 
 % \eea
 % \be
%&\Downarrow & \\&
 %\left( \nabla_M\,\hat{K}_0^{M}\right)^{\rm (sing)}&=0
 \eea
where $\nabla$ is the covariant derivative with respect to the five dimensional 
metric on the codimension 
one brane. From the previous formula, we learn that both the regular and singular 
parts must vanish simultaneously. However, it can happen that the covariant derivative 
induces singular contributions when it is applied to certain components of $\hat{K}_N^M$
by taking derivatives of $sign$ functions. This is indeed what happens in our case.
Consider the case $N=0$. The singular part of the previous formula tells us that
\be
   \frac{\partial}{\partial t}\,\left( K^{w_1}_{w_1}
   \right)^{sing} \,=\,\left(\nabla_M\,K^{M}_0 \right)^{sing},
\ee
where the left hand side contains the singular term 
associated with $K_{w_1 w_1}|_{sing}$, 
 while 
in the right hand side the singular terms come from the covariant derivatives. 
But the piece in the left hand side corresponds precisely to the term associated
with the six dimensional contribution at the intersection. Thus the six dimensional
contribution to the Friedmann equation on the intersection does not satisfy the 
energy conservation and there is an exchange of energy from codimension two brane 
to the higher dimensional space. 
   
In the light of this fact, one expects that the conservation of energy at the 
intersection does not hold. Instead, one finds the continuity equation
\be\label{nonconeq}
   \dot{\rho} +3 H \left( \rho+p\right)\,=\, 
   4 M_6^4\,
\,\frac{\partial}{ \partial \,\tau}\,\arctan{\left(\frac{\tan{\alpha}}{\sqrt{1-\dot{ z}_2^2}}\right)}.
\ee
Hence, the conservation of energy is ensured only when $\alpha$ vanishes, or when 
$\dot{z}_2$ is constant.
 
We close this section by summarising the equations that 
govern cosmology of the codimension one branes and the intersection;
\begin{eqnarray}
\frac{\Lambda_B}{10} &=& \bar{H}^2 -k_1^2 -k_2^2, \\
\lambda_1 &=& 6 M_5^2 (\bar{H}^2-k_2^2) + 8 M_6^4 k_1,\\
\rho_2 &=& -8 M_6^4 \frac{1}{\sqrt{1-\cos^2 \alpha \dot{z}_2^2}}
\left(k_1 \sin \alpha -(k_2 + \bar{H} \dot{z}_2) \cos \alpha \right),\\
\label{cod1}
\chi &=& 2 M_6^4 \frac{\dot{z}_2 \ddot{z}_2 \sin \alpha \cos \alpha}
{A(0,t) (1-\cos \alpha \dot{z}_2^2)^{3/2}}, \\
\rho &=& 3 M_4^2 H^2 +6M_{5,1}^3 \frac{k_2
+\bar{H} \dot{z}_2
}{\sqrt{1-\dot{z}_2^2}}+ 4 M_6^4\arctan{\left(\frac{\tan{\alpha}}{\sqrt{1-\dot{ z}_2^2}}\right)}
,
\label{intersection}
\end{eqnarray}
where $A(0,t)=1/(1+\bar{H} t + k_2 z_2(t))$, $\Lambda_B$ is the bulk 
cosmological constant, $\rho_2$ is the energy density on the moving codimension
one brane at $w_1=0$, $\chi$ is $(w_1,t)$-component of energy momentum tensor on the 
moving codimension one brane at $w_1=0$ and 
$\rho$ is the energy density at the intersection. 
The Hubble parameter at the intersection is given by Eq.~(\ref{genHin}).
The energy conservation at the intersection is given by 
Eq.~(\ref{nonconeq}). In the following we put $M_{1,5}=M_5$.
  
\section{Cosmological solutions} \label{sec:cosmsol}
In this section, we discuss the property of the 
cosmological solutions by focusing on the Friedmann 
equations on the moving four brane and at the 
intersection. 

\subsection{The branes at a right angle}
We first consider the simplest case in which the branes 
are at a right angle $\alpha=0$ and $\bar{H}=0$.
In this case, the energy momentum tensor on the moving 
four brane becomes the perfect fluid and there is no 
energy flow $\chi=0$. Using the cosmic time $\tau$, 
the 5D metric is given by
\be
d s_5^2\,=\,-d \tau^2+A^2(w_1, \tau)\, \left( d \vec{x}_3^2 +
d w_1^2 \right)\label{metr4br}, \quad A(w_1,\tau)= 
\frac{1}{1+k_1 w_1 + k_2 z_2(\tau)}.
\ee
Although the scale factor depends on $w_1$, the Hubble parameter in terms 
of the cosmic time is independent of $w_1$ and given by Eq.~(\ref{genHin}).
By expressing $\dot{z}_2$ in terms of the Hubble parameter $H$, we get 
\begin{equation}
\rho_2 = 8 M_6^4 \sqrt{H^2+k_2^2}.
\label{cod1-right}
\end{equation}
Since there is no energy flow, the energy density is conserved 
\be
\partial_{\tau} \rho_2 + 4 H (\rho_2 + p_2)=0.
\ee 
At the intersection, the Friedmann equation is given by
\be
\rho =3 M_4^2 H^2 + 6 M_{5}^3 k_2 \sqrt{1+\frac{H^2}{k_2^2}},
\label{intersection-right}
\ee 
and the standard continuity equation holds since 
$\tan \alpha=0$. 
Notice that the static brane gives a contribution of the 5D 
DGP form.

Since the Hubble parameters are equal in both the equations (\ref{cod1-right})
and (\ref{intersection-right}), by expressing $H$ as a function 
of the energies in the two cases and equalling the results, 
they will imply a fine tuning relation between the two homogeneous energy 
densities $\rho_2$ and $\rho$. Geometrically, this is because when the 
codimension one brane $\Sigma_2$ moves through the static bulk, it completely 
controls the dynamics of the brane $\Sigma_{\cap}$ that sits at the intersection 
with $\Sigma_1$. Then, $\Sigma_{\cap}$ can only follow the motion of $\Sigma_2$, 
without an independent dynamics on its own. The problem becomes clearer 
by the fact that the energy density and the Hubble parameter on the moving 
codimension brane do not depend on the coordinate $w_1$. Then, the 
energy density at the intersection actually fixes {\it all} the properties of 
the energy density on the moving brane $\Sigma_2$, including its equation 
of state.
   
There is a simple way out of a part of this problem. The fine-tuning we found 
is so strong because we demand that the moving brane $\Sigma_2$ keeps a 
straight shape -- that is, it cannot deform along the $z_1$ direction.  
Suppose however that we allow the moving codimension one brane to be free to 
deform and bend, forming a non-trivial angle $\alpha$ with $\Sigma_1$
which explicitly depends both on  $z_1$ and $t$. Then, the energy density on 
$\Sigma_2$ will explicitly depend on $z_1$. This implies that, although the energy 
density at the intersection must equal the energy density on $\Sigma_2$ calculated 
at $z_1=0$, nevertheless this fine-tuning is ameliorated with respect to the 
previous case. Indeed, it involves only the quantities calculated at the intersection. 
  
\subsection{Arbitrary angle between the branes}
Now let us consider the case in which $\alpha \neq 0$. In this case, the Hubble 
parameter at $w_1=0$ is given by Eq.~(\ref{genHin}). 
%However, it is important to 
%emphasise that this is only the value of the Hubble parameter when evaluated at 
%the position of the intersection: when calculated away from this point, it receives additional contributions coming from time derivatives of the angle $\alpha$, and it 
%becomes an inhomogeneous quantity. This is a crucial difference with respect to the 
%example studied in the previous subsection. 
 Again taking $\bar{H}=0$ for simplicity, 
the Friedmann equation on the moving four brane at the position of the intersection 
is given by
\bea
&&\hskip-.5cm\rho_2 = 8 M_6^4 (k_2\cos\alpha-k_1\sin\alpha) \sqrt{\frac{{ H}^2
    +k_2^2}{{H}^2\sin^2\alpha +k_2^2}}.
\label{frcod1angl}
\eea
On the other hand, the Friedmann equation at the intersection is given by
\bea\label{HubtdangleH}
 \rho &=&
3 M_4^2 H^2
%\nonumber\\ &+& 
+6 M_5^3 \,
%\left[ \frac{k_2}{\sqrt{H^2 \sin^{2}{\alpha}+k_2^2}}\left(k_1 \cos{\alpha}+k_2 \sin{\alpha}
%\left( 1+\frac{H^2}{k_2^2}\right)\right)
%+
k_2
\, \sqrt{1+\frac{H^2}{k_2^2}}
%\right]\nonumber\\
%&+&
+ \,4 M_6^4\,\arctan{\left[
\tan{\alpha}
\sqrt{1+ \frac{H^2}{k_2^2}}
\right]}~.
\eea  
When the angle $\alpha$ vanishes and $k_2$ remains finite,
we recover the results of the previous subsection. 
Notice also that comparison between Eqs.~(\ref{frcod1angl}) 
and (\ref{HubtdangleH}) imposes a fine-tuning
relation between $\rho$ and the energy density
$\rho_2$ of the codimension one brane, when evaluated at the intersection.
Nevertheless, this fine-tuning is much milder than the one we met in the 
previous subsection. This fine-tuning is associated with the restrictive Ansatz 
we have chosen for the bulk metric. 
 
The form of the previous Friedmann equation is quite complicated 
to study with full generality.  While the second term on the right 
hand side of Eq.~(\ref{HubtdangleH}) corresponds
to the well-known DGP-like term, the last term in the right hand side 
of Eq.~(\ref{HubtdangleH}) is less standard, 
and is associated with six dimensional contributions.

Notice that, in the limit of large $H$, this term approaches
the constant value $4 M_6^4\,sign{(k_2)}$, and then
can help to drive acceleration in this regime. We
will discuss in the last part of the paper more general situations,
where six dimensional contributions can provide sources
of acceleration for the induced cosmology at the intersection.

%This term is interesting
%because they contain $H$ at the denominator in a peculiar way,
%with interesting consequences for cosmology. The fact that $H$ appears at the 
%denominator implies that for large $H$ this term is suppressed. One may be 
%tempted to interpret this behaviour, at least partially, as a {\it relativistic effect}. Indeed, large $H$ means that the brane speed is approaching the speed of light (see the definition in formula (\ref{genHin})). But at higher and higher speeds, due to the 
%Lorentz contraction, an observer on the moving brane sees the intersection angle with the 
%static brane approaching the value $\pi/2$. But we know that in this limit the six 
%dimensional contributions at the intersection vanish, explaining why for large $H$
%this term is suppressed.

\smallskip

The continuity equation is given by  
\be
   \dot{\rho} +3 H \left( \rho+p\right)\,=\, 
   4 M_6^4\,
   \frac{\partial}{ \partial \,\tau}
\,
   \arctan{\left[
\tan{\alpha}
\sqrt{1+\frac{H^2}{k_2^2}} 
\right]},
  \ee
so, under the assumption that $k_2$ is non zero, the conservation of energy 
is ensured only when $\alpha$ vanishes, or when $H$ is constant.
For $\alpha \neq 0$, it is also necessary to have the energy flow 
on the moving codimension one brane, $\chi$, which also breaks the conservation 
of energy on the moving codimension one brane. Then we can understand that the 
energy at the intersection is transmitted to the moving codimension one brane.

%We can now return to the comment on the fine-tuning issues
%between brane energy densities. We learned that
%if there are fine-tunings between energy densities on the codimension one branes
%and the intersection, they involve only quantities evaluated at the intersection 
%itself. Indeed, the energy momentum tensor on the codimension one 
%brane is inhomogeneous, and it can take the preferred value, apart from the 
%intersection where it must be related to the one of the codimension two brane.
  
%  One can 
%  imagine that   the fact that there is an exchange of 
%  energy between codimension one and codimension
%  two branes, may also {\it help} to reduce the fine tunings.
%  Suppose indeed that we start with a static configuration, and
%  we add some energy momentum tensor at the intersection.
%  The branes start to move, and some energy density is consequently 
%  transmitted to the codimension one branes, due 
%  to (\ref{nonconeq}). In 
%  determinate situations,  this transmitted energy can 
%  contribute to satisfy the fine-tuning relations, and 
%  in special cases it may even automatically fulfil
%  them. 
%It would be interesting to check these ideas in 
%concrete cosmological models.
\section{Applications}\label{sec:cosmapp}
In this section, we derive some interesting consequences of the cosmological 
equations we discussed in the previous section. In the first two subsections, we examine 
some of the cosmological properties of the solutions we discussed in \cite{ckt1} in the present context. In the last subsection, we will instead derive a new selfaccelerating configuration in which the codimension one branes are {\it not} maximally symmetric, 
a case that we did not discuss in our previous work.  

\subsection{Self-tuning solutions}
Here we re-examine the selftuning solution
presented in \cite{ckt1}. 
We take $k_1=0$, $k_2=0$, $M_{5,1}=0$ and $\dot{z}=0$.
Then we have 
\begin{equation}
\lambda_4 = 3 M_4^2 \bar{H}^2 + 4 M_6^4  \alpha, 
\quad \bar{H}^2 =\frac{\Lambda_B}{10}.
\end{equation}
where $H=\bar{H}$. Then the expansion rate does not 
depend on $\lambda_4$. The self-tuning mechanism 
consists on the fact that if we change tension $\lambda_4$, $\alpha$
changes so that induced cosmology remains the same. Unfortunately, 
our embedding is not well suited to study the self-tuning property of the solution 
as $\alpha=\,$constant is actually
 imposed by hand.  A possible way out would be to consider a situation in which $z''(t,0) \neq 0$ at the intersection. This would generalise (\ref{con2impose}) with new pieces that would not necessarily impose that $\dot{\alpha}(t,0)=0$. It would be nice to study in more detail this kind
of generalisation to understand whether it can be compatible with the reflection 
symmetries of our system or not. It is important to understand whether the 
eventual self-tuning property would be compatible with the recovery of small 
scale 4d general relativity on the intersection, in order not to contradict big 
bang nucleosyntesis and other cosmological tests. In order for this last tricky 
issue to be solved, it seems to be necessary that the dynamical angle reacts 
{\em only} to the vacuum energy density component of the localised matter on the intersection: a priori this is rather counterintuitive. However, a few observations 
are in order here: it is well known that 6d brane worlds with conical singularities 
treat tension-type of matter on a completely different footing with respect to a 
generic fluid ($\omega\neq -1$)~\cite{puretension1}. In fact, also in the present 
setup the six-dimensional contribution to the 4d stress tensor
% (cfr. eq. (\ref{eninint})) 
has a tension-like structure; moreover as we showed, a generic fluid localised on 
the intersection does not seem to render the bulk geometry singular as opposed to what happens in the thin conical setups~\cite{puretension1}, but violates the 
conservation of energy. It is therefore not excluded that the self-tuning might be 
at work here.

\subsection{Self-accelerating solutions with $\bar{H} \neq 0$}
 Here we re-consider the self-accelerating solution
presented in \cite{ckt1}, for maximally
symmetric configurations.  
We assume there is no cosmological constant nor    
  matter in the 
system $\Lambda_B=\rho=\rho_2=\lambda_1=0$ with 
$\dot{z}=0$. Then
we get 
\begin{eqnarray}
0 &=& \bar{H}^2 - k_1^2 - k_2^2,\\
0 &=& 3 M_4^2 \bar{H}^2 + 6 M_5^3 k_2 + 4 M_6^4 
 \alpha, \\
0 &=& k_1 \sin \alpha - k_2 \cos \alpha,\\
0 &=& 6 M_5^3 (\bar{H}^2 -k_2^2) + 8 M_6^4 k_1.
\end{eqnarray}
If $k_1 <0$ and $k_2<0$, there are non-trivial 
solutions for $k_1$, $k_2$, $\alpha$ and $\bar{H}$.
The solution is roughly given by
\begin{equation}\label{roughsol}
\bar{H} \sim \frac{M_6^2}{M_4}, \quad M_5^3 \sim M_4 M_6^2.
\end{equation}
in accordance with what we found in the previous paper. 
Once $\alpha$, $k_2$ and $\bar{H}$ are fixed, 
Eqs.~(\ref{nonconeq}) and (\ref{intersection}) 
determine the cosmological dynamics with $\rho$ without 
ambiguity. The resulting cosmology is complicated due to the 
non-conservation of energy. Instead of dealing with this 
complicated case, we will discuss a simpler,
different  situation in the next subsection.  

\subsection{Self-accelerating solution with $\bar{H}=0$}

We consider the case $\bar{H}=0$, $k_1=0$, and $k_2 =  -\epsilon_2\beta \,|\sin{\alpha}|\, 
M_6$, for some positive constant $\beta$ and $\epsilon_2=\pm 1$.
To make a more direct comparison to analog studies
in the codimension one case,
 we focus on a regime in  which the quantity $H^2$ satisfies
the condition
\be
\label{finappro}
H^2 \gg
\beta^2 \sin^2{\alpha} M_6^2,\ee
and we will later
check  in which cases
 this relation can be satisfied in our context. 
The continuity equation becomes 
\be\label{nonconeqsec}
   \dot{\rho} +3 H \left( \rho+p\right)\,=\, 
   -4\,\epsilon_1\, \,M_6^4\,
   %}{\cos{\alpha}}\,
   \frac{\partial}{\partial \tau}\,\arctan{ \left[
   \frac{H}{\beta\,M_6\,\cos{\alpha}}\right]
}\,,
\ee
with $\epsilon_1\equiv -\frac{\sin\a}{|\sin\a|}\,=\,\pm 1$.
Notice that $\epsilon_1 =1$  means that we  are taking
negative values for the angle $\alpha$: in the conical
case, it would  correspond
to take an
excess angle for the conical singularity.

On the other hand
the Friedmann equation acquires the  form  
\be
  \rho \,\simeq\,3 \,M_4^2\,H^2
   - 6 M_5^3\,\epsilon_2\,H
 -
 4\,\epsilon_1\, \,M_6^4\,
   %}{\cos{\alpha}}\,
  \arctan{ \left[
   \frac{H}{\beta\,M_6\,\cos{\alpha}}\right]
}\,,
\ee
with $\epsilon_2 \,=\,\pm 1$
corresponding to the usual DGP 
choice of branches for the codimension one brane. 
The first term in the right hand side 
is dominant at large $H$, and ensures the correct four dimensional form 
for early time cosmology. The second term is the typical DGP
contribution, while the third term, a six dimensional effect, is 
less standard as we discussed before. The previous expression
can be easily rewritten as
\be
\frac{\rho}{3M_4^2} \,=\,
\left( H- \epsilon_2\, \frac{M_5^3}{M_4^2}
\right)^2
-\,\frac{M_5^6}{M_4^4}
-\frac{4\,
\epsilon_1 \,M_6^4}{3 M_4^2}
   %}{\cos{\alpha}}\,
  \arctan{ \left[
   \frac{H}{\beta\,M_6\,\cos{\alpha}}\right]
}\,,
\ee
from which we obtain   
\be
    H \,=\, \epsilon_2 \frac{M_5^3}{M_4^2}
    +\sqrt{ \frac{\rho}{3 M_4^2}
    + \frac{M_5^6}{M_4^4}+\epsilon_1\,\frac{4\,
  \,M_6^4}{3 M_4^2}
   %}{\cos{\alpha}}\,
  \arctan{ \left[
   \frac{H}{\beta\,M_6\,\cos{\alpha}}\right]
    }},
\ee
by imposing that the quantity inside the square root is positive.
Now, the choice $\epsilon_1=\epsilon_2=1$ corresponds to the standard DGP self-accelerating
branch, and the quantity inside the square root is always positive. It implies
that, even when $\rho$ vanishes, the Hubble parameter satisfies
the inequality
\be
H \ge  \frac{M_5^3}{M_4^2} \left(
1+\sqrt{1+ 
\frac{4\,
  \,M_6^4\,M_4^2}{3 M_5^6}
   %}{\cos{\alpha}}\,
  \arctan{ \left[
   \frac{H}{\beta\,M_6\,\cos{\alpha}}\right]}}
  \right),
\ee
and so we find a lower bound for $H$, as in the well-known 
self-accelerating branch of DGP model. 
    
This case is very similar to the standard five dimensional
case, since the acceleration is mainly driven by the effects of the 
codimension one brane. It is however also possible to study the case in 
which $M_5=0$, to understand whether six dimensional effects provide 
acceleration by themselves. We will focus 
on this case, choosing $\epsilon_1=1$,
 for the remaining discussion. 
Then, the continuity 
equation (\ref{nonconeqsec}) can be formally integrated as
\be
\rho\,=\,\rho_0\,\left( \frac{a(t)}{a_0}\right)^{-3(1+\omega)}
\,\exp\left[
-\int_{X(\infty)}^{X(H)}\,\frac{d \tilde{x}}{\frac{3 M_4^2}{4 M_6^2}
\beta^2 \cos^2{\alpha}\,\tan^2{\tilde{x}}-\tilde{x}
}
\right],\ee
defining\be
X(H)\equiv\arctan{
\left[\frac{H}{\beta
\cos{\alpha}\,M_6}\right]},
%\,\left( 1- \frac{4 \beta M_6^5}{3\,H^3\, M_4^2\cos{\alpha}}
%\right)^{\frac13},
\ee
and introducing
the constants $\rho_0$, $a_0$  
(corresponding to the quantities 
evaluated at a fiducial time) and calling
 $w=p/\rho$. 
This solution shows that, in the limit $H \to \infty$, one recovers the 
usual relation between energy density and scale factor
because in this limit the last factor in the previous
equation becomes $1$.
Using the Friedmann equation,
% Plugging this expression 
%in the Friedmann equation, a little manipulation leads to the following relation
%\be
%\rho_0^{\frac32}\,\left( \frac{a(t)}{a_0}\right)^{-\frac92 (1+\omega)}
%\,=\, \left(3 M_4^2\right)^{\frac32} \left( H^3\,-\,
%\frac{4 \beta M_6^5}{3\, M_4^2\cos{\alpha}}\right),
%\ee
%that is the differential relation that our scale factor 
%must satisfy, for a given equation of state. 
%We notice that from this one gets
 it is straightforward
 to get the following relation for the acceleration:
\be
\frac{\ddot{a}}{a}\,=\,\dot{H}+H^2\,=\,
-\frac{1}{2 M_4^2}\,\left[
\frac13+\omega
-\frac{8M_6^4}{3}\,
\frac{\arctan{\left[
H/\left(\beta
\cos{\alpha}\,M_6\right)\right]}
}{3 M_4^2\,H^2-
 4 \,M_6^4\,
   %}{\cos{\alpha}}\,
  \arctan{ 
  \left[
H/\left(\beta
\cos{\alpha}\,M_6\right)\right]
  }}
\right]\,\rho,
%\left[2-3\,\left(1+\omega
%\right)\left(1
%-\frac{4 \beta M_6^5}{3\,H^3\, M_4^2\cos{\alpha}}\right)\right]
%\,\frac{H^2}{2},
\ee
so we have the acceleration when
\be
\omega \,<\, 
-\frac13+
\,\frac{8M_6^4}{3}\,
\frac{X(H)
}{3 M_4^2\,H^2-
 4 \beta\,M_6^4\,
   %}{\cos{\alpha}}\,
X(H)
  }
.
\ee
Then for small $H$ we learn that the six dimensional 
contributions help to provide the acceleration and it 
can be achieved even when $\omega >  -\frac13$.

To conclude, we discuss the late time cosmological
evolution for our system. 
In an expanding universe,
at late times $\rho \to 0$;
 then $H$ approaches a constant value, given
by the solution of the equation
\be
H^2-\frac{4 M_6^4}{3 M_4^2}\, \arctan{ 
  \left[\frac{
H}{\beta
\cos{\alpha}\,M_6}\right]} 
\,=\,0.
\label{fineq}
\ee
Let us study two limiting cases, compatible
with the condition (\ref{finappro}),
in which equation (\ref{fineq})
admits simple solutions.
The first is the case in which $\alpha$
is not too small, from which applying (\ref{finappro})
in  (\ref{fineq}) we get the equation
\be\label{sac1}
H^2\,\simeq \frac{4 M_6^4}{3 M_4^2}\,.
\ee
In this case, $H$ can be rendered small by choosing 
$M_4$ sufficiently larger than $M_6$. On the other
hand, condition (\ref{finappro}) requires $M_6 \gg \beta\,M_4$,
so the constant $\beta$ must be chosen correspondingly
 small.
 Notice that  relation (\ref{sac1})  is similar to the one we already
met in eq. (\ref{roughsol}), 
although in the present case we have $M_5=0$.

The second possibility is to consider $\alpha$
extremely small, and at the same time 
focus on a regime in which $H \ll \beta M_6$. 
The the solution of  (\ref{fineq})  provides
a different relation between $H$ and the mass scales:
\be
H \simeq \frac{4 M_6^3}{3\,\beta\, M_4^2}\,,
\ee
and also in this  case we can get a sufficiently small value
of $H$ by choosing $M_4$ much larger than $M_6$.

%This cosmological model can be studied along the lines of the
%analysis of \cite{dvaliturner}. 
Notice that the present examples of 
self-acceleration require a different analysis,  
with respect to the ones discussed
in the previous subsection originally found in \cite{ckt1}. This  
is because, for the particular choice of our embedding,  
the codimension one branes do not need to be maximally symmetric, nor empty. 
It would be nice to understand whether in this case ghosts are present
in the low energy spectrum, and if so how do they manifest themselves.

\section{Conclusions and Open Issues}\label{sec:conc}
In this paper, we explored the cosmological features of a codimension two 
brane with induced gravity terms, sitting at the intersection between two 
codimension one branes in six dimensions.
We found that the cosmological expansion at the intersection is controlled by 
contributions coming from the codimension one branes, and from the six dimensional 
bulk. We first showed that the effect of the codimension one branes on the Friedmann
equation at the intersection have the well-known DGP form. Then, we learned that 
six dimensional contributions are much less standard. They can have an important role
for late time cosmology  providing a new source of the geometrical acceleration, 
controlled by the angle between the branes. 
At the same time, they are also associated with a violation of the energy 
conservation at the intersection, allowing a flow of energy between the 
codimension two brane and the higher dimensional space. We discussed consistency 
relations that matter on the codimension two brane must satisfy and the 
connection with the choice of energy momentum tensor localised on the codimension 
one branes. 

The main aim of this work was to formulate a general and powerful formalism based
on the approach of {\it mirage cosmology} that can be used to study 
cosmological solutions in this and similar models, and to apply it to a couple 
of representative examples. Due to the fact that the codimension one branes
that intersect with general angle are not homogeneous and isotropic,
a numerical analysis is likely to be needed in order to analyse
in full details the cosmological evolution of this class of models.    
As a natural continuation of the present work, it would be interesting to study the low
energy effective action for the light modes  associated with our brane configurations. 
This analysis would be necessary in order to investigate whether ghosts are present
in the spectrum of the low energy theory, and, if so, whether they can be eliminated 
with a mechanism similar to the one of \cite{newdvali}. We leave these issues to 
future work.
 
\subsection*{$\hskip6.8cm$Acknowledgements}
OC is partly supported by a Marco Polo fellowship of the University of Bologna.
KK is supported by STFC. GT is supported by MEC and FEDER under grant FPA2006-05485, 
by CAM under grant HEPHACOS P-ESP-00346, and by the UniverseNet network  (MRTN-CT-2006-035863). 
OC and GT are grateful to the Institute of Cosmology and 
Gravitation at the University of Portsmouth for support and warm hospitality while this work was initiated. 

%\newpage
 \bigskip

%\footnotesize
%\begin{multicols}{2}

%\end{multicols}
\end{document}